\documentclass[3p,times,procedia,twocolumn,letterpaper]{elsarticle}
\usepackage{times}
\usepackage{amssymb}
\usepackage{amsmath}
\usepackage{graphicx}
\usepackage{float}
\usepackage{booktabs}
\usepackage{subfig}
\usepackage{epstopdf}
\usepackage{tabularx}
\usepackage[linesnumbered]{algorithm2e}
\usepackage{fancyvrb}
\usepackage{amsthm}
\usepackage{color}
\usepackage{epstopdf} 
\usepackage[ ps2pdf-options={-dPDFSETTINGS=/prepress} ]{pstool}

\usepackage{amssymb}
\usepackage[figuresright]{rotating}

\begin{document}

\begin{frontmatter}

\title{Thread-Based Obfuscation through Control-Flow Mangling}
%\tnoteref{Threaded-Based Control-Flow Obfuscation}}




\author{Rasha Salah Omar\fnref{fn1}}
\ead{rasha.omar@ejust.edu.eg}

\author{Ahmed El-Mahdy\fnref{fn2}}
\ead{ahmed.elmahdy@ejust.edu.eg}

\fntext[fn1]{Computer Science and Engineering Department, Egypt-Japan University of Science and Technology, Egypt, http://www.ejust.edu.eg}

\author{Erven Rohou\fnref{fn3}}
\ead{erven.rohou@inria.fr}

\fntext[fn2]{Computer Science and Engineering Department, Egypt-Japan University of Science and Technology, and on-leave from Alexandria University, Egypt, http://www.ejust.edu.eg}

\fntext[fn3]{Inria -- Centre de Recherche Inria Rennes - Bretagne Atlantique, http://www.inria.fr}

%% e.g. \dochead{17th International Conference on Dynamical Processes in Excited States of Solids}

\begin{abstract}

The increasing use of cloud computing and remote execution have made program security especially important. Code obfuscation has been proposed to make the understanding of programs more complicated to attackers. In this paper, we exploit multi-core processing to substantially increase the complexity of programs, making reverse engineering more complicated. We propose a novel method that automatically partitions any serial thread into an arbitrary number of parallel threads, at the basic-block level. The method generates new control-flow graphs, preserving the blocks' serial successor relations and guaranteeing that one basic-block is active at a time using guards. The method generates $m^n$ different combinations for $m$ threads and $n$ basic-blocks, significantly complicating the execution state. We provide a correctness proof for the algorithm and implement the algorithm in the LLVM compilation framework.
\end{abstract}


\end{frontmatter}



\section{Introduction}

With the advent of cloud computing, software security becomes especially important~\cite{a}. In particular, software security researchers have been concerned to evaluate the methods that protect software systems against reverse engineering threats. Those can be exploited by software hackers to discover the software vulnerabilities and inject malicious code. 

One practical security approach is to use software \emph{obfuscation}; it is a software security
mechanism that transforms the original program into a functionally-equivalent counterpart~\cite{b,d}; the obfuscated program has the same semantics as the original, but it is much more complex to understand by reverse engineers~\cite{w}.

Parallelism and multi-threading have been proposed to increase performance. Parallel programs are notoriously difficult to debug and reason about, and for that reason parallelism is a nice ingredient for
obfuscation. 

This paper proposes a new obfuscation technique based on control-flow restructuring and multi-threading. The method allows for arbitrary distribution of basic-blocks to an arbitrary number of threads. Thus for $m$ threads and $n$ basic-blocks, there are exactly $m^n$ different distributions. The method uses guards to transfer execution from one thread to another, thus guaranteeing sequential execution semantics. To some extent, these guards can be seen as the key to rebuilding the original code, and they can be obfuscated by standard methods, such as splitting each code block into multiple blocks (only according to its data dependence). Furthermore, dummy processes could be inserted to the parallel code. Our method has the following advantages:

\begin{itemize}
\item it makes static analysis more complex -- if not impossible -- as the actual control-flow is mangled and hidden;
\item it allows for generation of different versions of the same program (software diversity), thereby increasing the resilience of versions to attacks;
\item the method is generally independent from other control-flow obfuscation methods, and thereby it can be combined for increased complexity.
%Rewrite last item
\end{itemize}

This paper is organised as follows: Section~\ref{related} reviews related control-flow obfuscation methods. Section~\ref{method} describes our thread-based obfuscation method, illustrating it with a simple example, and provides a general correctness proof for the method. Section~\ref{discussion} discusses the obfuscation and performance implications of the method. Finally, Section~\ref{conclusion} concludes the paper.

\section{Related Work}
\label{related}

According to Collberg et
al.~\cite{d}, software obfuscation is classified into three
categories. The first one, operating at source code level, consists in lexical transformations, shuffling the code identifiers and getting rid of the comments and
debugging information. The second category applies data transformations, with the main focus on data layout change. Finally, the
third category is the control-flow transformations, aiming at making the control-flow unintelligible from the attackers~\cite{d,f}. The technique we propose fits in this latter category.

\subsection{Parallel Control-Flow Obfuscation}
Control-flow obfuscation is a method to obscure the actual
control-flow of the code, without changing the semantics of the
original code. Control-flow transformation itself is divided into three
subcategories. \emph{Aggregation} breaks up logically dependent portions
of code and merges the independent portions. \emph{Ordering} alters
the order of computations of the code. Finally, \emph{Computations}
inserts dummy code or make obscure changes to the source by inserting supplementary dummy computations so as to further confuse the attacker~\cite{d}.
%,z}.

The \emph{computation} transformation includes methods such as insertion of dead or irrelevant code, extension to loop conditions, and conversion of a reducible into a non-reducible flow graph. One of the most important methods to do this is to increase parallelism of the code by two methods. First, redundant non-profitable task can be created then a portion of code is paralleled with this task, so the hacker cannot deduce which thread will have the actual portion to be run. Second, a sequence of data dependent statements in a portion of code can be splitted, then they run in parallel. The control for the correct execution will be using synchronization primitives~\cite{e}. The latter is the closest to our method, however it splits simple, in-order sequence of instructions (without control-flow dependence) rather than general, complex, control-flow graphs as per our method.

Hsin-Yi Tsai, Yu-Lun Huang, and David wagner~\cite{r} proposed a quantitative analysis framework through using control-flow graphs after applying some obfuscation transformations. In their study, code parallelization makes it more difficult for reverse-engineers to comprehend the main purpose of the software. Moreover, there were two proposed solutions. First, splitting each code block into multiple blocks, but only according to its data dependence. Second, dummy processes could be inserted to the parallel code~\cite{d}. 
%Chandrasekharan \cite{r} reports that parallel execution cannot be expressed through using a simple graph representation. Our approach differs, as it is not constraint by data-dependence for creating parallel threads. 

\section{Embedding Control-Flow into Multiple Threads}
\label{method}
Our proposed method aims to obscure code running in hostile environments, such as cloud systems, by
dividing its actual single thread control-flow graph into multiple
control-flow graphs which run separately in threads. The real semantics
of the program is guaranteed by using synchronization
primitives. These primitives consists of a guard for each basic-block
in each control-flow that waits until it is set to be executed. For
$m$ threads and $n$ basic-blocks, the number of combinations for this
algorithm reaches $m^n$ that provides more security. The algorithm
works on basic-block level and it confirms that only one basic-block
runs at a time and in its original order.

We explain in this section all specifics of the proposed approach. Starting from the algorithm main steps followed by its correctness proof and finally an example to clearly elucidate our work.

\subsection{Our Obfuscation Algorithm}
In our proposed algorithm we have four main steps. The first step decomposes the base control-flow graph basic-blocks, randomly, among a given number of sets. The second  step computes the immediate successors for each basic-block in the same set. Note that the original successor of a block may be moved to a different thread, and control returns to the former thread only at a later block. The third step inserts guards before each basic-block to prevent it from running before its actual serial execution sequence. Finally, the fourth step  generates corresponding, new, parallel functions (threads), each containing one of the basic-block sets. We also ensure that each function contains only one entry and end basic-block. 

We use the Informal Compiler Algorithm Notation (ICAN) format~\cite{j} for describing our algorithm. The first and fourth steps are generally straightforward. We therefore focus here on steps two and three.
 
Step 2 is concerned with computing immediate successors; the number of successors varies from 1 to many, depending on the distribution of the basic-blocks.

%\linesnumbered
\begin{algorithm}
{\bfseries procedure} GetImmediateSuccessors (BBcur, BBset, BaseCFG)

BBcur :{\bfseries  in} BBType
 
BBset : {\bfseries in} BBsetType

BaseCFG : {\bfseries in} CFGType

Result : {\bfseries out} BBsetType

{\bfseries begin} 

BB : BBsetType

BB $:=$ \{ BBcur \}
\label{temp_ini}

seenBefore : BBsetType

seenBefore $:=$ $\phi$ 

\nl\While {BB $\neq \phi$}  { 
\label{l0}

\ $I$ $:=$ baseImmediateSuccessor(BB , BaseCFG) 
\label{l1}

\ Result $\cup$ $:=$ $I$ $\cap$ BBset
\label{l2} 
 
\ mayNext $:=$ $I$ $-$ BBset $\cap$ $I$
\label{l3}

\ BB $:=$ mayNext $-$ seenBefore
\label{l4}

\ seenBefore $\cup :=$ mayNext
\label{l5}
}
{\bfseries end}

\caption{GetImmediateSuccessor}
\label{GetImmediateSuccessor}
\end{algorithm}

The procedure of \emph{GetImmdediateSuccessor}, shown in Algorithm \ref{GetImmediateSuccessor}, is used to determine the new successors in each function for each basic-block. It is called on each basic-block of the original control-flow graph, BaseCFG. It computes the immediate successor set, Result, for an input basic-block set, BBcur, a basic-block partition set, BBSet, and the original control-flow graph, BaseCFG.

Generally, an immediate successor can belong to other basic-block partitions; however, an immediate successor of which can, recursively, belong to the current set. Therefore, the algorithm keeps track of all such blocks in a temporary set, BB, and iteratively adds new such blocks. Initially, the BB is intialised to BBcur (line \ref{temp_ini}), iteratively seen blocks are removed.

The line \ref{l0} iterates till  BB is empty. It becomes empty when we exhaustively visit all possible immediate successors blocks (that belong to the current partition, BBset). In line \ref{l1}, we get the immediate successor of all BB elements and store into the set $I$. In line \ref{l2}, we intersect the current partition, BBset, with $I$ to partial immediate successors, storing them into Result. In line \ref{l3}, we compute possible predecessors by computing the difference between $I$ and the intersection between $I$ and BBset (currently obtained immediate successor) and store the result into mayNext. In line \ref{l4}, we filter the already visited predecessors, seenBefore, and update BB. Finally, in line \ref{l5}, we append the already visited predecessors into the seenBefore to avoid visiting in later iterations.

The execution of each basic-block is guarded by a flag associated with it. After finishing the execution of a basic-block, the current next successor block's flag is set to 1. The control-flow then switches to the correct successor depending on the values of flags (ie, the one with the flag set to 1). We, therefore, need to insert a switching basic-block for that purpose. In particular, we insert two new basic-blocks; the first is a $Wait$ block, that waits for any successor flag to be 1. The next block performs the switching part, \emph{SwitchBasicBlock}.

The variables used in the original program and the flags introduced as guards have to be made volatile, as well as, global to be visible for all newly created functions. Furthermore, we should care that these global variables are read in right way.

%The value of each memory access should be right and when it is written from one processor, it should be invalidated or updated in all other processors.
Hardware cache coherence will guarantee that corresponding cache lines are properly invalidated and values updated at readers' locations. This mechanism has a cost, which may be worsened in case of false sharing. 
 To minimize it, we force a large alignment for every global variable.
%In particular, large alignment is inserted to avoid false sharing.

\subsection{Proof of Correctness}

The proposed algorithm was proved in three main points.

Firstly, every basic-block in the new partitions maps to one and only one of the old basic-blocks. According to each iteration, there is only one basic-block that is set into only one partition. Every partition takes each basic-block from the $BaseCFG$ for once.

Secondly, the immediate successor could be reached from each basic-block and every partition should be waiting for the immediate successor of the current basic-block. Initially all threads are waiting on the dominant set of basic-blocks that belong to its partition and reachable from the $BaseCFG$ entry node. After executing a basic-block, the thread waits on all dominant set basic-blocks that belong to its partition following execution of the executable basic-block. Therefore all threads are always waiting on the current dominant set of basic-blocks in its partition. Then, the dominant set is updated after executing each basic-block in its partition. In other words, each thread is waiting on all its basic-blocks to define a dominant basic-block such that no basic-block should be executed before first execution of the dominant basic-block. The dominant set is determined according to the $BaseCFG$.

Finally, there is only one basic-block which can be active at a time. Each basic-block in each partition has a guard that determines when it could be active. According to the original sequence of instructions, before each basic-block being active, the corresponding guard should be set and once the basic-block finish its work, this guard should be reset and the next basic-block should be activated using the same criteria.

\subsection{Simple Example to Our Proposed Approach}
\label{example}
We illustre here with a simple example the effect of applying our algorithm on a (naive) program that computes prime numbers. The pseudo-code of the program is written as following:
%\begin{verbatim}
%function checkNumber
%begin
 % for counter in range(0, 1000000)
 % begin
 %   if (counter == 0)
 %     print "zero"
 %   elseif(counter % 2 == 0)
  %    print "even"
 %   else
 %     print "odd"
 % end
%end
%\end{verbatim}
\noindent
\small
\begin{verbatim}
function checkPrime
begin
  for counter in range(0, 100000)
  begin
    flag = 0
    for i in range (2, counter / 2)
    begin
      if (counter % i == 0)
      begin
      	flag = 1
      	break
      end
    end
    if(flag == 0)
      print "prime"
    else
      print "not a prime"
  end
end
end
 \end{verbatim}
\noindent
\normalsize

\begin{figure}[]
\centering
\includegraphics[height=8cm, width=6cm]{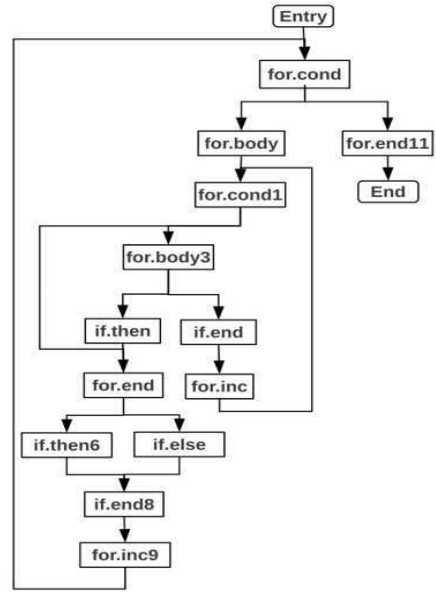}
\caption{The Original Program CFG in the Example}
\label{fig:eps1}
\end{figure}

The original serial program's control-flow graph is shown in Fig. \ref{fig:eps1}. The program consists of 16 basic-blocks, two loops, and a number of branches.

% Our technique task is to extract all basic-blocks in this program to do some processes to be obfuscated.
The obfuscating process computes the basic-blocks of the program and randomly divides them into a number of sets. In our example, we divided it into four sets, randomly. Furthermore, each set contains different random number of original basic-blocks. In addition, the edges are constructed to get the new control-flow graphs with new guards each time according to the new sets. 

\begin{figure}[]
\centering
\subfloat[Obfuscated Result:CFG1]
{
\includegraphics[height=5.5cm, width=2cm]{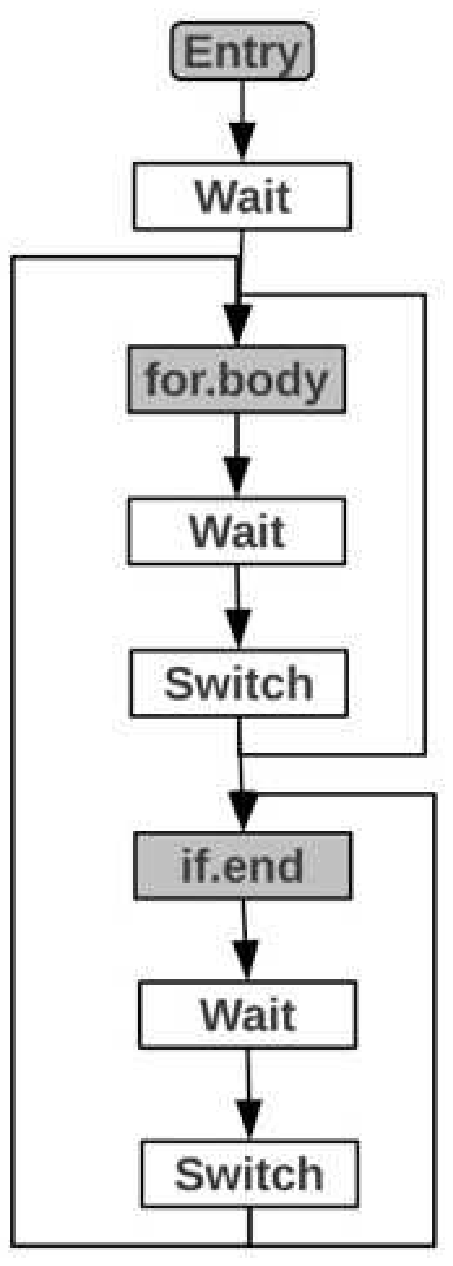}
%\caption{CFG1}
\label{fig:eps2}
}
%\end{figure}
%\begin{figure}[]
\centering
\subfloat[Obfuscated Result:CFG2]
{
\includegraphics[height=4.5cm, width=4cm]{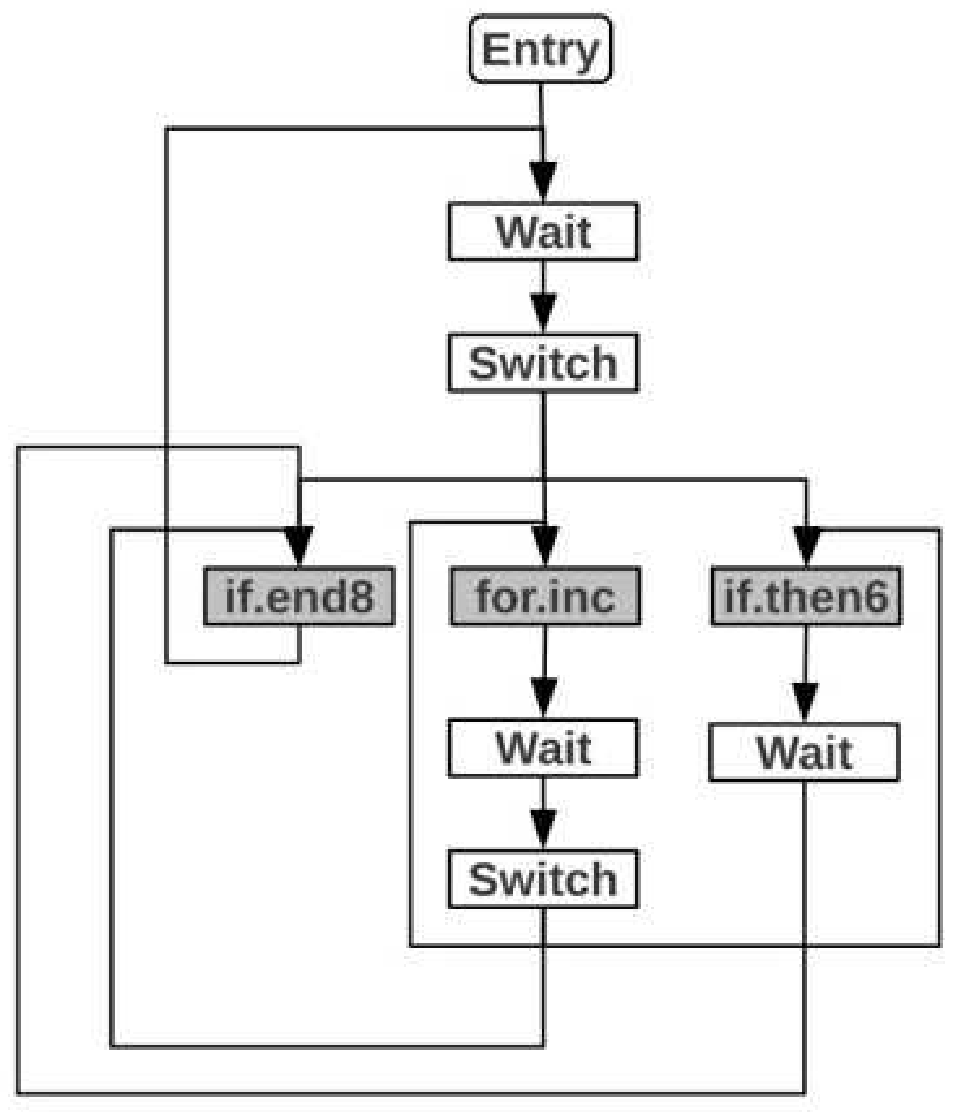}
%\caption{CFG2}
\label{fig:eps3}
}
%\end{figure}
%\begin{figure}[]
%\begin{figure}[]

\centering
\subfloat[Obfuscated Result:CFG3]
{
\includegraphics[height=7cm, width=6cm]{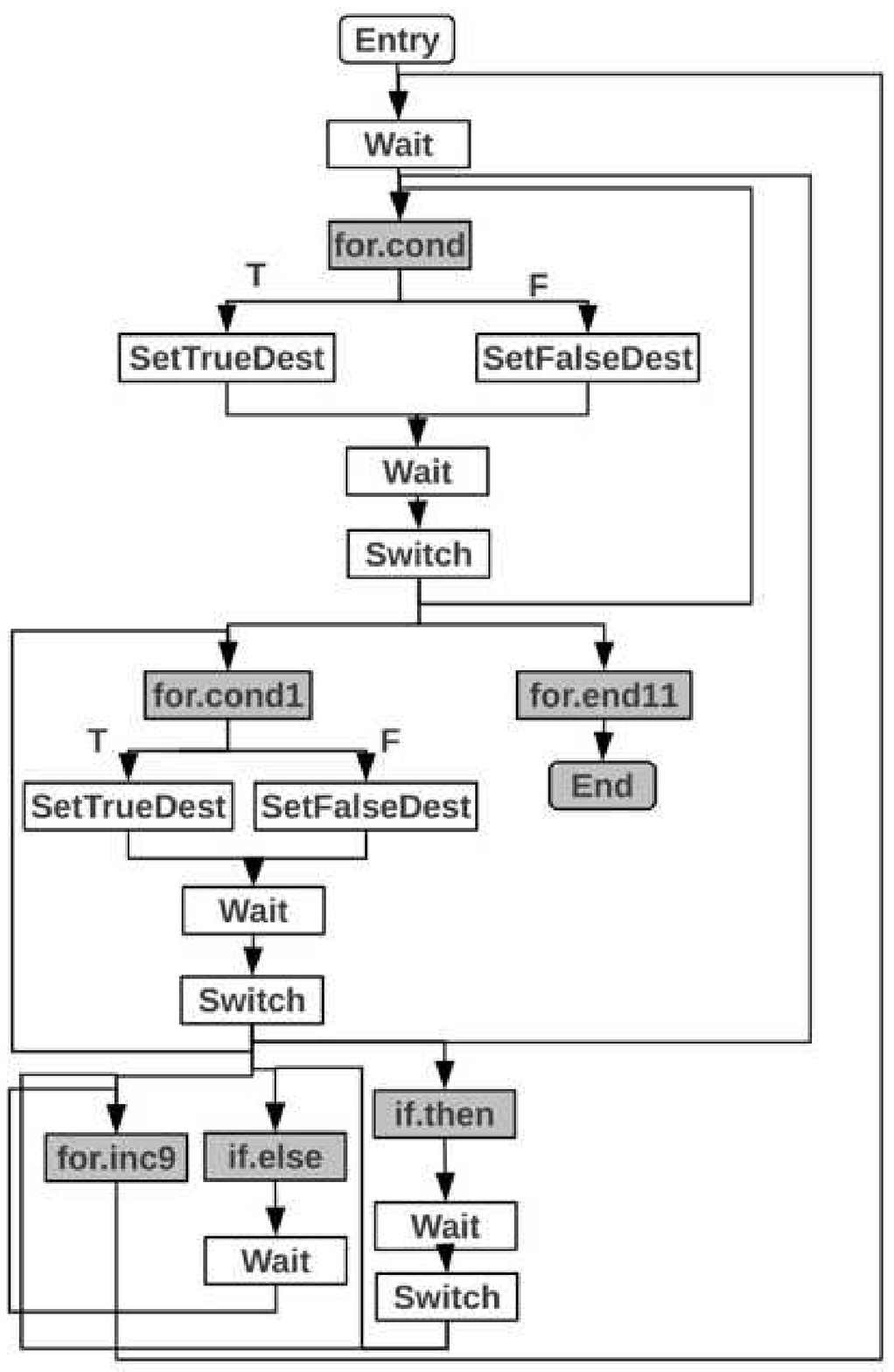}
%\caption{CFG3}
\label{fig:eps4}
}

\centering
\subfloat[Obfuscated Result:CFG4]
{
\includegraphics[height=6.5cm, width=6cm]{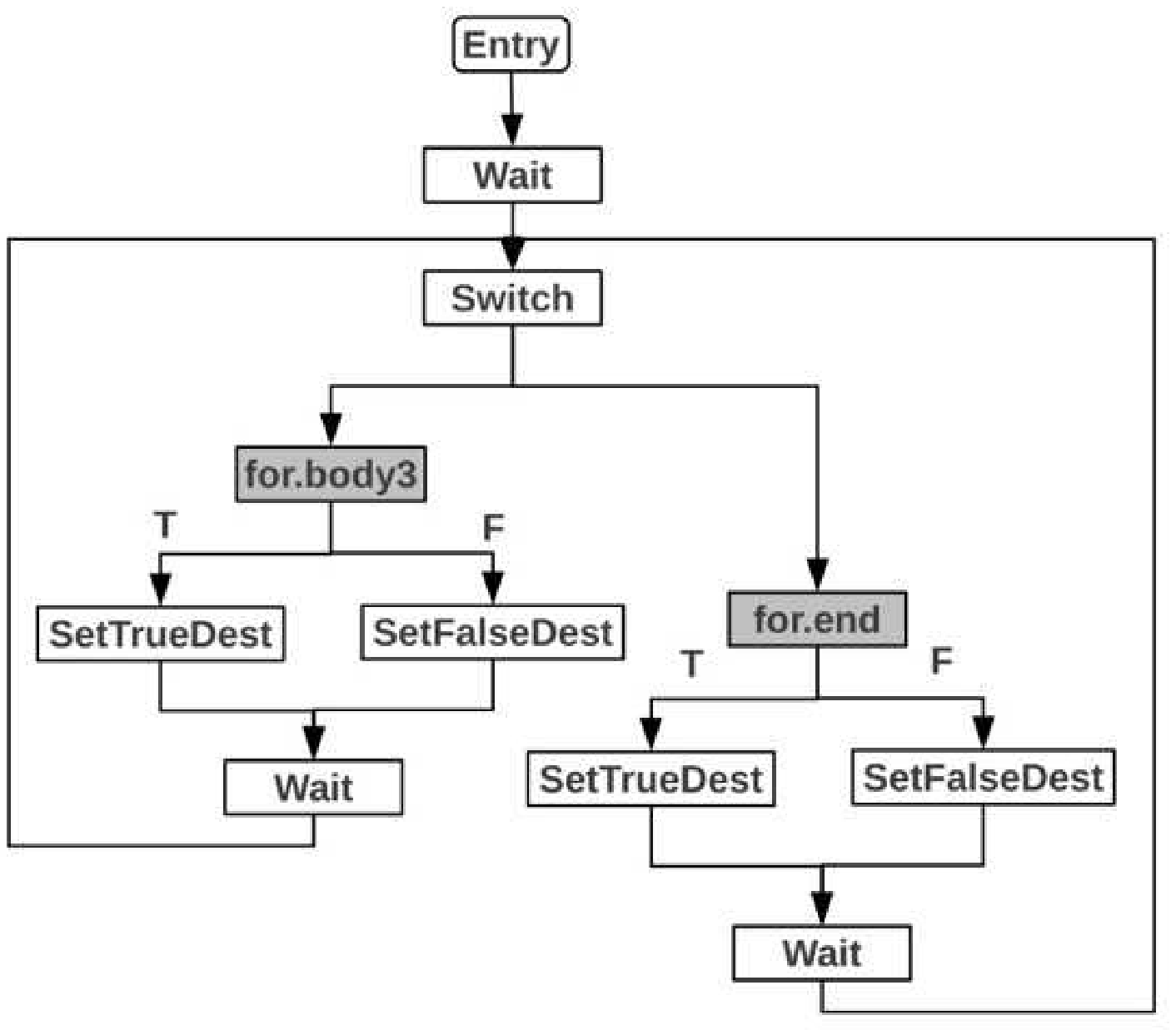}
%\caption{CFG4}
\label{fig:eps5}
}
\caption{One Example of Obfuscated CFGs}
\end{figure}

The result for the algorithm application to the program is shown in Fig. \ref{fig:eps2}, Fig. \ref{fig:eps3}, Fig. \ref{fig:eps4} and Fig. \ref{fig:eps5}. 

\subsection{Implementation}
To further validate the algorithm, we have implemented it on the LLVM compilation framework~\cite{llvm}. We added a new transformation pass that uses Pthreads to generate parallel threads. The system is used on a simple kernels to validate the correctness of the algorithm. The kernels included calculating even numbers, Fibonacci sequence, and prime numbers. The implementation is not optimised for performance, which is a subject for future work.

\section{Discussion}
\label{discussion}

\subsection{Resilience and Potency}
The key, what the attacker is looking for, is the distribution
order. This information can be reconstructed from writes into global
variables at the end of basic-blocks and active waits at the beginning
of blocks. Even though we did not implement it so far, additional
protection would hide these accesses, for example by accessing the
variables through pointers (the insight is that alias analysis is a
difficult problem), or by reusing the same location when we can prove
they are never used at the same time.

Static reverse engineering is therefore much more difficult, if not
impossible. Dynamic analysis would be similar in spirit to what one would do on the sequential code, but with many more threads need to be tracked. We could also use races to
make the analysis even more complicated, and confuse automatic tools.

Moreover, we increase the dimension space for an attacker as the method is highly independent on other obfuscation methods. Thereby the use of threading effectively increases the dimensionality of the space, resulting in an exponential increase in complexity.

\subsection{Cost}

According to the splitting of the program, the obfuscated program
is significantly slower than the original. Generally, the obfuscation effect on the program depends on
the communication via memory. Thus, performance degrades
and cost increases as well. 

The naive implementation incurs between one and two orders of magnitude slowdown. This is due to the cost of communication between threads (and possibly false sharing). Another cost is due to the use of global variables and the lack of register promotion at this time.

\subsection{Perspective}

One approach to hide the cost, is only obfuscate critical functions such as password checking function. However, for general intellectual property performance-critical functions general performance improvement is highly sought.

The main performance bottleneck is communicating values across threads. For threads residing on the same core, the wait spins on a shared flag, thereby increasing the memory demand. Careful spin-wait optimisations (such as the use of the \verb!pause! instruction in the x86 architecture) improves performance. For threads residing on different cores, the cost would include communicating across the cache hierarchy. Typically, such spin-waits are 10-20 times slower than L1 cache accesses, and therefore the choice of basic-blocks can be constraint by corresponding  performance thresholds. Thereby trading-off some obfuscation complexity to gain speed. Another possible optimization is to promote global variables into registers (register promotions) for specific code regions with no external modifications for shared variables. 

%Mitigate slowdown, and study how far we can be to original performance.

%Good to load balancing?

\section{Conclusion and Future Work}
\label{conclusion}

This paper proposes a novel method that mangles any control-flow graph into many threads; the mangling is done randomly, without data-flow or control-flow dependence constraints. The method exploits the complexity inherent in parallel programming and debugging to obfuscate program. The method is capable of generating  $m^n$ different decompositions on $m$ threads and $n$ basic-blocks (in the original control-flow graph). The main aim of the paper is to present a preliminary study of the proposed method, focusing on the correctness, and the performance degradation extend. The correctness is validated by a proof, as well as by testing a real LLVM-based implementation. Naive implementation is  performance bounded by the thread-to-thread communication through the cache hierarchy, resulting in up to 10-20 times slowdown.

Future work would focus on improving the performance throughout extending the liveness analysis across threads, thereby allowing for register allocations for some variables that are not communicated across threads. Moreover, from the point view of obfuscation, interesting future work would be to obfuscate multiple program threads at the same time; that would allow for a mangled concurrent execution state, resulting in further complexity.

\section{Acknowledgment}
This work is partially supported by the French Ministry of Foreign
Affairs and Egyptian Ministry of Higher Education and Research through the
IMHOTEP program, grant number 27467YJ. The work is also partially supported by a scholarship from the Egyptian Ministry of Higher Education and Research.

\bibliographystyle{elsarticle-num}
\bibliography{paper2}
\end{document}